\documentclass{article}
             \usepackage{graphicx}
            \usepackage{amsmath,amssymb}
            \begin{document}
            \title{Effects of gauge theory based number scaling on geometry }
            \author{Paul Benioff,\\
            Physics Division, Argonne National
            Laboratory,\\ Argonne, IL 60439, USA\footnote{E-mail:pbenioff@anl.gov}}
            \date{}
            \maketitle

            \begin{abstract}
             Effects of local availability of mathematics (LAM) and space time dependent number scaling on physics and, especially, geometry are described. LAM assumes separate mathematical systems as structures at each space time point.  Extension of gauge theories to include freedom of choice of scaling for number structures, and other structures based on numbers, results in a space time dependent scaling factor based on a scalar boson field.  Scaling has no effect on comparison of experimental results with one another or with theory computations. With LAM  all theory expressions are   elements of mathematics at some reference point. Changing the reference point introduces (external) scaling.  Theory expressions with integrals or derivatives over space or time include scaling factors (internal scaling) that cannot be removed by reference point change. Line elements and path lengths, as integrals over space and/or time, show the effect of scaling on geometry.  In one example, the scaling factor goes to $0$ as the time goes to $0$, the big bang time.   All path lengths, and values of physical quantities,  are crushed to $0$ as $t$ goes to $0.$  Other examples have spherically symmetric scaling factors about some point, $x.$ In one type,  a black scaling  hole, the scaling factor goes to infinity as the distance, $d$, between any point $y$ and $x$ goes to $0.$  For scaling white holes, the scaling factor goes to $0$ as $d$ goes to $0.$  For black scaling holes, path lengths from a reference point, $z$, to $y$ become infinite as $y$ approaches $x.$ For white holes, path lengths approach a value much less than the unscaled distance from $z$ to $x.$
            \end{abstract}


            \section{Introduction}
            As is well known, mathematics is very important to physics. The goal of theoretical physics is to construct mathematical models to explain the physical universe.  A good theory is a model which can successfully predict the outcomes of many different experiments on many different physical properties.  Yet it is not clear why mathematics should be so successful in constructing good theories.  There does not seem to be a basic a priori reason why the physical world should be so amenable to description by mathematical theories.

            This problem has bothered many others, including this author. Wigner's paper \cite{Wigner}, on the unreasonable effectiveness of mathematics in the natural sciences, engendered  other papers on the relationship between mathematics and physics \cite{Omnes,Plotnitsky,Hamming}. All this work recognizes how closely entwined physics and mathematics are. A paper by Tegmark  \cite{Tegmark} expresses this in an extreme fashion by claiming physics \emph{is} mathematics.

             One approach to this problem is to work towards  development of a coherent theory of physics and mathematics together \cite{BenCTPM1,BenCTPM2}.  Such a theory would be expected to treat physics and mathematics as one coherent whole rather than as two separate disciplines. Development of such a theory requires that one have some idea of what physics and mathematics are.

             There is an enormous literature on the nature and foundations of mathematics. Here these questions will be bypassed by assuming that mathematics includes the description and properties of  many  different mathematical systems. Here  mathematical  systems are defined \cite{Barwise,Keisler} as structures consisting of a base set, a few basic operations, relations, and constants.  For each system type the structures are required to satisfy axioms relevant to the system type. Examples are  structures for the rational numbers\begin{equation}\label{Ra}\overline{Ra}=\{Ra,\pm,\times,\div,<,0,1\} \end{equation} that satisfy the axioms for the smallest ordered field \cite{rational}, the complex numbers, \begin{equation}\label{Complex}\bar{C}=\{C,+,-, \times,\div, <, 0,1\},\end{equation} that satisfy, axioms for  an algebraically closed field of characteristic $0$ \cite{complex} and Hilbert spaces\begin{equation}\label{Hilbert}\bar{H}= \{H,\pm,\cdot, \langle-,-\rangle,\psi\}\end{equation} that satisfy axioms for a complex, normed, inner product space that is complete in a norm defined from the inner product, \cite{Kadison}.  Scalar vector multiplication is denoted by $\cdot$ and a generic vector is denoted by $\psi$.

             Here mathematics is considered to consist of a large collection of structures for the different types of mathematical systems.  For each type of system many structures are possible. Thesee are all part of mathematics. The  structures can differ in that the base sets can be different. Also how the basic operations and relations are defined can change.  The only requirement among the many structures of a given type is that structure $A$ satisfies the axioms for the type if and only if structure $B$ does.

             \section{Mathematics is Local}\label{ML}

             The connection to physics begins with the assumption that mathematics is local.  This assumption means that  a collection, $\bigcup_{x},$ of mathematical systems of different types is associated with each space time point $x$ of a manifold, $M$. In addition the mathematics available to an observer or intelligent being, $O_{x},$ is  assumed to be limited to that in $\bigcup_{x}.$

             This assumption is based on the idea that the mathematics available to $O_{x}$ is that which is stored in his brain. At any time $t$ the mathematics in an observers brain is a subset of that in $\bigcup_{x}.$ Mathematical information in a textbook or described in a seminar is not available to $O_{x}$ until the information has reached $O_{x}'s$ brain.

             In addition, to keep things simple, the assumption is made here that an observer, or more specifically his brain, can be localized at a point. This is clearly false as storage and use of a large amount of information requires a finite space time volume \cite{Lloyd}.

             Note that the the totality of mathematics available to an observer at $x$ is  limited to that within $\bigcup_{x}.$ It follows that $\bigcup_{x}$ must be large enough to include the mathematics used by physics.  It must also be such that $O_{x}$ can describe relations and maps between systems in $\bigcup_{x}$ and systems in $\bigcup_{y}$ at other points $y.$ Systems in $\bigcup_{x}$ are used for such descriptions.

             Maps between systems in $\bigcup_{y}$ and $\bigcup_{x}$  play an important role in this work.  They start with maps on the different types of numbers. These are then extended to other types of systems that include numbers in their description (such as vector spaces, algebras, group representation or any system that includes multiplication by scalars). These maps are isomorphisms between  structures at different points.  They are also parallel transforms \cite{Mack} in that they correspond to or define the notion of "same value as" between elements at different points.

             For example let $\bar{C}_{x}=\{C_{x},Op_{x}, 0_{x},1_{x}\}$ and $\bar{C}_{y}=\{C_{y},Op_{y}, 0_{y},1_{y}\}$ be complex number structures at $x$ and $y$ Define the map $F_{y,x}\bar{C}_{x}=\bar{C}_{y}$ by \begin{equation}\label{Fxy}\begin{array}{c}F_{y,x}c_{x} =c_{y} \mbox{ for all $c_{x}$ in $C_{x}$}\\\\F_{y,x}Op_{x}=Op_{y},\hspace{.5cm} F_{y,x}0_{x}=0_{y},\;F_{y,x}1_{x}=1_{y} \end{array}\end{equation} The element $c_{y}$ has the same value in $\bar{C}_{y}$ as $c_{x}$ has in $\bar{C}_{x}.$ $Op$ denotes the four field operations, $\pm,\times,\div.$

             For Hilbert spaces, $\bar{H}_{x}=\{H_{x},\pm_{x},\cdot_{x}, \langle-,-\rangle_{x},\psi_{x}\}$ and $\bar{H}_{y}=\{H_{y},\pm_{y},\cdot_{y}, \langle-,-\rangle_{y},\psi_{y}\}$, one has a similar isomorphic map $V_{y,x}\bar{H}_{x}=\bar{H}_{y}$ where \begin{equation}\label{Hxy}\begin{array}{c}V_{y,x}\psi_{x} =\psi_{y} \mbox{ for all $\psi_{x}$ in $H_{x}$},\hspace{.5cm}V_{y,x}\pm_{x}=\pm_{y}\\\\ V_{y,x}c_{x}\cdot_{x}\psi_{x}=F_{y,x}c_{x}V_{y,x} \cdot_{x}V_{y,x}\psi_{x}=c_{y}\cdot_{y}\psi_{y},\\\\V_{y,x}\langle\psi_{x},\phi_{x}\rangle_{x} =F_{y,x}(\langle\psi_{x},\phi_{x}\rangle_{x})= \langle\psi_{y},\phi_{y}\rangle_{y}.\end{array}\end{equation}

             \section{Scaling of Number structures}\label{SNS}

             The next step taken in the approach used here and in other work, \cite{BenSPIE2,BenNOVA,BenINTECH}, is to introduce scaling between mathematical systems at different points. Scaling can be done for each type of number system.  For example, let $\bar{R}_{y}$ and $\bar{R}_{x}$ be real number structures at points $y$ and $x$ and $r_{y,x}$  a positive real number in $\bar{R}_{x}.$ Define the scaled representation of $\bar{R}_{y}=\{R_{y},Op_{y},<_{y},0_{y},1_{y}\}$ on $\bar{R}_{x}=\{R_{x},Op_{x},<_{x},0_{x},1_{x}\}$ by the structure, \begin{equation}\label{Rrx}
             \bar{R}^{r}_{x}=\{R_{x},\pm_{x},\frac{\times_{x}}{r},r\div_{x},<_{x},0_{x},r_{x}\}.\end{equation} To save on notation, $r_{y,x}$ is denoted here by $r.$ The scaled structure can also be represented by \begin{equation}\label{Rxr}\bar{R}^{r}_{x}=\{R_{x},\pm^{r}_{x},\times^{r}_{x},\div^{r}_{x}, <^{r}_{x},0^{r}_{x},1^{r}_{x}\}.\end{equation}

             This shows that in the scaled structure, the number value, $1,$ corresponds to the number value, $r$ in $\bar{R}_{x},$ The  addition and subtraction operations are unchanged, multiplication in $\bar{R}_{y}$ corresponds to multiplication and division by $r$ in $\bar{R}_{x},$ division in $\bar{R}_{y}$ corresponds to division  and multiplication by $r$ in $\bar{R}_{x},$ and $<$ and $0$ are the same in  the scaled representation of $\bar{R}_{y}$ on $\bar{R}_{x}$ as they are in $\bar{R}_{x}.$ In general any number value $a_{y}$ in $\bar{R}_{y}$ corresponds to the number value $ra_{x}$ in $\bar{R}^{r}_{x}.$ Here $a_{x}$ is the same number value in $\bar{R}_{x}$ as $a_{y}$ is in $\bar{R}_{y}.$

             The definitions of the scaled operations, $Op^{r}_{x}$ in terms of those in $Op_{x}$ are not arbitrary.  They must be such that $\bar{R}^{r}_{x}$ satisfies the real number axioms, for a complete ordered field \cite{real}, if and only if $\bar{R}_{x}$ does. Eqs. \ref{Rrx} and \ref{Rxr} satisfy these conditions.

             The transformations between the structures, with scaling included, can  be defined by factoring $F_{y,x}$ into two isomorphisms as in $F_{y,x}=Z_{y,x}W_{y,x}$ where
             \begin{equation}\label{FZWR}F_{y,x}\bar{R}_{x}=Z_{y,x}W_{y,x}\bar{R}_{x}=Z_{y,x} \bar{R}^{r}_{x}=\bar{R}_{y}.\end{equation} Note that the usual case with no scaling is the special case here where $r=1.$ Then $W_{y,x}$ is the identity and $Z_{y,x}=F_{y,x}.$

             The introduction of scaling requires that one drop the association of elements of the base set to  specific number values.  For example, the above shows that the element of the base set $R_{x}$ that has value $r_{y,x}$ in $\bar{R}_{x}$ has value $1$ in $\bar{R}^{r}_{x}.$ This shows that base set elements have no intrinsic values.  The  elements of $R_{x}$ acquire values only within the structure  containing them.  They are different for different structures.

             The only exception is the element with value $0$ in a structure. This value is invariant under scaling transformations.  In this sense $0$ corresponds to the number vacuum.

             This description of scaling extends to other types of numbers.  For complex numbers it can be taken over directly. Here the scaled representation, $\bar{C}^{r}_{x}$ of $\bar{C}_{y}$ on $\bar{C}_{x}$ is given by \begin{equation}\label{Crx}\bar{C}^{r}_{x}=\{C_{x},\pm_{x},\frac{\times_{x}} {r},r\div_{x}, 0, r_{x}\}.\end{equation} Scaling applies to rational numbers provided they are regarded as subsets of real or complex numbers,  The same holds for integers and natural numbers.

             Scaling affects all systems that include scalars in their structures (such as closure under multiplication by scalars). Vector spaces, algebras, group representations are examples of these systems. For example the scaled representation, $\bar{H}^{r}_{x},$ of the Hilbert space $\bar{H}_{y}$ on $\bar{H}_{x}$ is given by \begin{equation}\label{Hrx}\bar{H}^{r}_{x}= \{H_{x},\pm_{x},\frac{\cdot_{x}}{r},\frac{\langle r\phi,r\psi\rangle_{x}}{r},r\psi\}. \end{equation} More details on scaling are given in \cite{BenSPIE2,BenNOVA,BenINTECH}.

             \subsection{Description of $r_{y,x}$}
             The space time dependence of $r_{y,x}$ is conveniently described by a  real vector field $\vec{A}(x).$ Let $y=x+\hat{\mu}dx$ be a neighbor point of $x.$ Define $r_{y,x}$ by
             \begin{equation}\label{ryxn}r_{y,x}=e^{\vec{A}(x)\cdot\hat{\mu}dx}.\end{equation} If $y$ is distant from $x$, define $r_{y,x}$ by \begin{equation}\label{rpxy}r^{p}_{y,x}=\exp(\int_{0}^{1} \vec{A}((p(s))\cdot\frac{dp}{ds}ds)=\exp(\int_{p}\vec{A}dp).\end{equation} Here $p$ is a path from $x$ to $y$ parameterized by $s$  with $p(0)=x$ and $p(1)=y.$ The path dependence is shown by the superscript, $p,$ on $r_{y,x}.$ $\int_{p}$ is a path integral along $p.$

             If $\vec{A}(x)$ is the gradient of a scalar field, $\Theta,$ as in $\vec{A}(x)=\nabla_{x}\Theta,$ then $r_{y,x}$ is path independent and Eqs. \ref{ryxn} and \ref{rpxy} become \begin{equation}\label{ryxnT}r_{y,x}=e^{\nabla_{x}\Theta\cdot\hat{\mu}dx}\end{equation}and \begin{equation}\label{rTxy}r_{y,x}=e^{\Theta(y)_{x}-\Theta(x)}.\end{equation} Here $\Theta(y)_{x}=F_{x,y}\Theta(y)$ is the same value in $\bar{R}_{x}$ as $\Theta(y)$ is in $\bar{R}_{y}.$ Transfer to a common structure is necessary because subtraction is defined only within a structure, not between structures.

             \section{Effects of Scaling on Physics}\label{ESP}
             The effect of scaling on physics is based on the observation that any comparisons of, or operations on numbers, that are associated with two different space time locations, include the effects of scaling.  The reason is that comparisons or operations are not defined for numbers in different structures at different locations.  They are defined only within structures. This means that any theoretical physics expressions that involve integrals or derivatives over space, time, or space time must include scaling as these operations require use of number values or other mathematical elements in different structures  at different locations.

             An example of the effect of scaling and locality of mathematics is the replacement of the usual expression of a wave packet $\psi=\int d^{3}z |z\rangle\langle z|\langle z|\psi\rangle$ by \begin{equation}\label{psixSc}\psi_{x}=\int_{x} d^{3}z_{x}e^{\Theta(z)_{x}-\Theta(x)} |z\rangle_{x}\langle z|\langle z|\psi\rangle_{x}.\end{equation}The subscript $x$ means that the integral is evaluated in mathematical systems at $x.$  Each vector value of the integrand is transferred, with scaling, to a vector in $\bar{H}_{x}.$ Here, and in much of the following, the scaling factor will be assumed to be defined by a scalar field, $\Theta$ as is done in Eqs. \ref{ryxnT} and \ref{rTxy}.

              The problem with the usual expression for a wave packet is that, with  local availability of mathematics,  the integral in $\psi=\int d^{3}z |z\rangle\langle z|\langle z|\psi\rangle,$ as the limit of a sum of vectors for different space locations, is not defined. The reason is that the integrands belong to different Hilbert spaces, with $|z\rangle\langle z|\langle z|\psi\rangle$ a vector in $\bar{H}_{z}.$ Linear superposition is defined only within Hilbert space structures, not between different structures.

              This problem can be fixed by use of parallel transformation operators, $F{z,x},$ without scaling present. The inclusion of scaling, shown in Eq. \ref{psixSc}, follows from the additional assumption of the freedom of choice of number structures, and structures based on numbers, at different space and time locations. This is similar to the freedom of choice of bases in vector spaces as used in gauge theories \cite{Yang,Montvay}.

             Inclusion of scaling also affects the momentum operator in quantum mechanics. The usual expression $\vec{p}\psi =(\hbar/i)(d/d\vec{x})\psi(\vec{x})$ is replaced by the canonical momentum \begin{equation}\label{momSc}\vec{p}_{x}\psi_{x} =\frac{\hbar}{i} (\frac{ d}{d\vec{x}}+\vec{A}(x)) \psi(\vec{x}).\end{equation} Here $\vec{A}(x)$ is the gradient field of $\Theta.$

             Path integrals also show the effect of scaling. The usual expression for the time evolution of a quantum state $\psi(t)$ from $t_{i}$ to $t_{f}$ can be expressed by \begin{equation}\label{psitfti} \psi(t_{f},y)=\int\langle y|e^{-iH(t_{f}-t_{i})}|x\rangle\langle x|\psi(t_{i})\rangle dx.\end{equation} Here $\psi(t_{f},y)$ is the amplitude for finding a system in state, $\psi,$ at location, $y$ at time $t_{f}.$  The path integral expression for the matrix element, $\langle y|e^{iH(t_{f}-t_{i})}|x\rangle =K(y,t_{f};x,t_{i}),$, is \cite{Rothe}\begin{equation}\label{Kyx} K(y,t_{f};x,t_{i})=\int d\gamma e^{\frac{i} {\hbar}S(\gamma)}\end{equation} where \begin{equation}\label{Sg}S(\gamma)= \int_{t_{i}}^{t_{f}}\mathcal{L}(\gamma(t), \dot{\gamma})dt.\end{equation} The integral $\int d\gamma$ is over all paths $\gamma(t)$ where $\gamma(t_{i})=x$ and $\gamma(t_{f})=y.$

             Local availability of mathematics with scaling affects these equations in two ways.  One is the transfer of the endpoints of the paths $\gamma$ at  $y$ as a number value triple in $\bar{R}^{3}_{y}$ and $x$ as a number value triple in $\bar{R}^{3}_{x}$ to scaled number value triples in $\bar{R}^{3}_{x_{0}}.$  The other is the transfer of the $S(\gamma)$ integrands at each point $\gamma(t)$ as number values in $\bar{R}_{\gamma(t)}$ to scaled number values in $\bar{R}_{x_{0}}.$  The result is seen in the replacement of Eq. \ref{psitfti} by the scaled representation of  $\psi(t_{f},y)$ at $x_{0}$ as  \begin{equation}\label{psitftiSc}\begin{array}{l} \psi(t_{f},y)^{Sc}_{{x}_{0}}= \\\\\hspace{.2cm}e^{\Theta(y)_{x_{0}}-\Theta(x_{0})} \int_{x_{0}}e^{\Theta(x)_{x_{0}}-\Theta(x_{0})}(\int_{x_{0}} d\gamma e^{\frac{i} {\hbar}S(\gamma)^{Sc}_{x_{0}}})\langle x|\psi(t_{i})\rangle dx_{x_{0}}.\end{array}\end{equation} The scaling factors transfer, with scaling, the endpoints of the paths, $\gamma,$ to number values in $\bar{R}_{{x}_{0}}.$

              The scaled action is given by\begin{equation}\label{SSc}S(\gamma)^{Sc}_{x_{0}}= \int_{x_{0},t_{i}}^{t_{f}}e^{\Theta(\gamma(t)) -\Theta(x_{0})}\mathcal{L}(\gamma(t), D_{t}\gamma)dt.\end{equation} The replacement of $\dot{\gamma}$ in the Lagrangian by $D_{t}\gamma$ where, \begin{equation}\label{Dtc}D_{t}\gamma =[\frac{d}{dt}+\vec{A}(\gamma(t))] \gamma,\end{equation} takes account of the effect of scaling on the time derivative of a path.

             Scaling also affects the variation of $S(\gamma)$ to obtain the equations of motion in that extra terms involving the derivatives of $\Theta$ appear in the Euler Lagrange equations.  Some details are given in \cite{BenNOVA}.

             The field, $\Theta,$ also appears in gauge theories in that its presence results in the expansion of the gauge group $U(n)=U(1)\times SU(n)$ for each $n$ to $GL(1,C)\times SU(n)$ \cite{BenSPIE2,BenNOVA,BenINTECH}. For each point $x,$ a field $\psi$ has a value, $\psi(x),$  in the vector space $\bar{V}_{x}$ \cite{Montvay}. If one replaces the common scalar field $\bar{C}$ for all the $\bar{V}_{x}$ with $\bar{C}_{x}$ for each $x$ and introduces the scaling field $\Theta$, then the representations of the spaces $\bar{V}_{y}$ at $x$ correspond to scaled structures $\bar{V}^{r_{y,x}}_{x},$ as shown in Eq. \ref{Hrx} for Hilbert spaces. Use of the usual derivation of the covariant derivative and restriction of the Lagrangians to terms invariant under local $U(1)$ transformations \cite{Cheng} gives the result that $\Theta$ is a scalar boson field for which mass is optional.  Additional interaction terms of the form $g\vec{A}(x)\psi(x)$ appear in the Lagrangians. Here $g$ is a coupling constant and $\vec{A}(x)=\nabla_{x}\Theta.$

             In field theory the action is given by \begin{equation}\label{Sp}S(\psi)=\int\mathcal{L} (\psi,D\psi)dx.\end{equation} With scaling and the local availability of mathematics included, the action becomes\begin{equation}\label{SScpsi}S(\psi)^{Sc}_{x_{0}}=\int_{x_{0}}e^{\Theta(x)_{x_{0}}- \Theta(x_{0})}\mathcal{L}(\psi,D^{Sc}\psi)dx_{x_{0}}.\end{equation} Here $D\psi$ is the usual covariant derivative and $D^{Sc}\psi=(D+\vec{A}(x))\psi$ accounts for the effect of scaling on the derivative.

             Unlike the case for $S(\gamma)^{Sc}_{x_{0}}$ in quantum mechanics, the presence of the scaling factor in the integral of Eq. \ref{SScpsi} has no effect on the variation of $S(\psi)^{Sc}_{x_{0}}$ with respect to $\psi.$ The resulting equations of motion, with $D^{Sc}$ replacing $D,$ are the same as those obtained with no exponential scaling factors present.

             \section{Restrictions on $\Theta$}\label{RT}

             These examples  show that the presence of number scaling, with local availability of mathematics, affects theoretical predictions in physics.  So far, there is no experimental evidence of  the presence of number scaling. This places restrictions on the value of $\vec{A}(x)$ and the coupling constant $g$.  In particular, the great accuracy of quantum electrodynamics without $\vec{A}$ shows that the ratio of $g$ to the fine structure constant must be very small.

             The absence of experimental evidence for scaling  in measurements of properties of quantum mechanical systems implies that for all  locations $x,y$ for  which amplitudes $\psi(x)$ and $\psi(y)$ are not negligible, one must have \begin{equation}\label{Tyx0}\Theta(y)_{x}-\Theta(x) \simeq 0\end{equation} to within experimental error. The size of the region within which this equation holds is determined by the fact that quantum states, $\psi$, at least those that are prepared and measured by observers, have negligible amplitudes outside a region that is small on a laboratory scale. It follows that the integral over all space, as in Eq. \ref{psixSc}, can be restricted to a small region.

             Here the restriction will be expanded in that space and time integrals for quantum states\footnote{This excludes quantum states of the whole universe and probably multiverse states.} will be restricted to a much larger region, $L,$ that is occupiable by us as observers.
             This region should be large enough to include all locations we may occupy in the future.  It should also include locations of other intelligent beings on distant planets with which we can communicate and discuss physics and mathematics.

             These restrictions suggest that $L$ should include the solar system.  It should also include nearby stars.  A reasonable estimation of the size of $L$ is that it should be a sphere of a few lightyears in diameter that is  centered on the solar system. The exact size is not important.  The only restriction is that it be a small fraction of the cosmological universe\footnote{It has been noted that the size of the space region in which we can hope to discover if intelligent beings even exist, let alone communicate with them, is a sphere of about $2,500$ lightyears in diameter \cite{AMSci}.}

             It follows that, for all points $x,y$ in $L,$ $\vec{A}(x)\simeq 0$ and $\Theta(y)_{x}-\Theta(x)\simeq 0.$ However, this restriction does not apply for points outside $L$.  Thus $\vec{A}(x)\neq 0$  is possible for most of the universe.  Also the fact that scaling depends  on $\nabla_{x}\Theta$ and on differences only between values of $\Theta$ means that one is free to choose a value of $\Theta$ at some point without affecting scaling.  Here the value $\Theta(x)=0$ will be chosen for a point $x$ in $L$. Then \begin{equation}\label{Ty0} \Theta(y)\simeq 0\end{equation}for all $y$ in $L$.

             It is also the case that scaling plays no role in  comparisons among outputs of computations and experimental measurements.   This is the case even if scaling is not negligible in $L$ and the computations and experiments are done in $L$. The reason is that the output of a computation or measurement at $y$ is not a real number in $\bar{R}_{y}.$ It is a physical system in a state $\alpha_{y}$ that is interpreted as a number in $\bar{R}_{y}$.  If $I_{y}$ is the interpretative map at $y$, then $I_{y}(\alpha_{y})$ denotes the numerical output  at $y$.

             Assume that another computation or measurement at $x$ gives an output state $\beta_{x}$.  The number value associated with this is given by $I_{x}(\beta_{x})$ as a number in $\bar{R}_{x}.$

             Comparison of outputs of these two operations, one at $x$ and one at $y$ does not involve comparison of the number values, $I_{y}(\alpha_{y})$ and  $I_{x}(\beta_{x}).$ Such a comparison would include a scaling factor between $\bar{R}_{y}$ and $\bar{R}_{x}.$  Instead the comparison is done by physically transmitting the states $\alpha_{y}$ and $\beta_{x}$ (or the relevant information contained in the states) to a common point $z$ where the outputs can be compared locally. If $T_{z,x}$ and $T_{z,y}$ denote the transmissions, then comparison of the outputs of these two operations consists of a comparison of the number values,  $I_{z}(T_{z,y}(\alpha_{y}))$ and $I_{z}(T_{z,x}(\beta_{x})).$ Since both numbers are in $\bar{R}_{z},$ no scaling is needed.

             These considerations suggest that, if number scaling plays a role in physics or geometry, then its role is limited to cosmological aspects of both physics and geometry.  Any influence of number scaling on physics and geometry is, at most, limited to events and properties of systems that are far away in either space or time or in space time.\footnote{It is worth pointing out that cosmological observations are all limited to local observations, in $L$ of incoming photons or particles.  One can include operations with measurements on these incoming signals, but the operations are all local. One cannot go far away to prepare these signals or to influence them.}

             \section{Effects of Scaling on Geometry}\label{ESG}

             As has been shown elsewhere \cite{BenNOVA}, local availability of mathematics and number scaling have an  effect on geometry.  This is a consequence of the assignment of separate real number structures, $\bar{R}_{x}$, to each point, $x,$ of $M$ instead of just one structure, $\bar{R}$, common to all $x.$ Affected aspects of geometry include coordinate systems, line elements, and path lengths. These are discussed here as examples of the effect of local availability of mathematics and number scaling on geometry.

             \subsection{Coordinate systems}\label{CS}

              A consequence of the local availability of mathematics is that for each point, $x,$ the coordinate system $CS_{x}$ associated with  a point, $x,$ in  an $n$ dimensional manifold, $M$, is described by  real number tuples in $\bar{R}_{x}^{n}.$  The origin of $CS_{x}$ can be anywhere.  It does not have to be at $x.$ Let $y$ be another point in $M$ with associated coordinate system $CS_{y}$. The relation between the descriptions of a point in $CS_{y}$ and in $CS_{x}$ may seem problematic because they use number value tuples in different real number structures.

              In the absence of scaling this is not a problem.  To see this let $x$ be the origin of $CS_{x}.$  Let $CS_{y}$ be a coordinate system with point locations given as tuples of number values in $\bar{R}^{n}_{y}.$  A point $z$ is described in $CS_{y}$ by a number tuple, $a^{n}_{y}$ in $\bar{R}^{n}_{y}.$ It is described in $CS_{x}$ by a number tuple, $a^{n}_{x}$  in $\bar{R}^{n}_{x}.$

              The parallel transformation operator, $F_{x,y},$ is an isomorphism of $\bar{R}_{y}$ onto $\bar{R}_{x}$ that preserves values.  It follows that \begin{equation}\label{Rnx}
              \bar{R}^{n}_{x}=(F_{x,y}\bar{R}_{y})^{n}.\end{equation} The corresponding mapped coordinate system, $(CS_{y})_{x}=(F_{x,y}\bar{R}_{y})^{n}$ is identical to $CS_{x}$ because the  parallel transform maps preserve the notion of same value. That is,  $a^{n}_{x}=(F_{x,y}a_{y})^{n}$ is the same number value tuple in $\bar{R}^{n}_{x}$ as $a^{n}_{y}$ is in $\bar{R}^{n}_{y}.$

              It follows that, with no scaling present, changing the tuples of real number structures used to describe points in coordinate systems  has no effect on the values of point representations as number tuples. The coordinate systems are entirely equivalent. They all lie on top of one another with parallel axes. The coordinates of a point $y$ in any coordinate system $CS_{z}$ are the same as they are in $CS_{x}$, The parallel transform change of $\bar{R}_{z}$ to $\bar{R}_{x}$ does not change the coordinate number values. Note that the change described consists of the change of reference number tuples, $\bar{R}^{n}_{y}$ to $\bar{R}^{n}_{x}$.  It is completely different from an actual translation  or rotation of a coordinate system.

\begin{figure}[h!]\begin{center}\vspace{1cm}
            \rotatebox{270}{\resizebox{110pt}{110pt}{\includegraphics[150pt,350pt]
            [500pt,700pt]{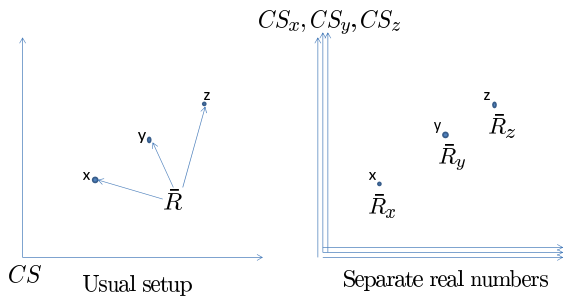}}}\end{center}\caption{Coordinate Systems for the usual setup with one real number structure for all space points and with separate structures for each point. The righthand panel shows the coordinates systems all lying on top of one another with parallel axes. This is a consequence of the number preserving property of the parallel transform maps.}\label{SPIE31}\end{figure}

             Figure \ref{SPIE31} shows the effect on coordinate systems of separate number structures at each point with no scaling. The left panel shows the  usual coordinate system description  with one real number structure for all space points. The right panel shows the coordinate system descriptions with separate real number structures at each point for a two dimensional space. Three of the infinitely many coordinate systems, one for each point, are shown.

              Since $(CS_{y})_{x}$ and $(CS_{z})_{x}$ are identical to $CS_{x},$ it follows that the parallel transform maps convert  $CS_{x}$ into a globally valid coordinate system, at least for Euclidean and Minkowski spaces. This shows that assigning separate number structures to each space point, with no scaling, is equivalent to the usual setup with just one universal number structure for all points.  The only difference is that the conversion to the usual setup is relative to some reference point, $x.$

             \subsection{Scaling present}\label{SP}

             Scaling changes the picture in that it is not possible to extend locally valid coordinate systems to globally valid ones, even for Euclidean and Minkowski geometries. The problem is that coordinate values in $CS_{y}$ are scaled relative to those in $CS_{x}$ and the scaling factor depends on $y.$ This corresponds to a situation different from that in Figure \ref{SPIE31} in that  the representations, $(CS_{y})_{x}$ and $(CS_{z})_{x}$, of $CS_{y}$ and $CS_{z}$ on $CS_{x}$ are scaled by factors that depend on $y$ and $z.$ However, the coordinate systems all have the same origin relative to a reference coordinate system as the $n$ tuple $(0_{z})^{n}$ is an invariant under parallel transformation and scaling.

             Nevertheless, it is possible to describe the effects scaling has on geometry.  A point $y$ with coordinate values, $a^{n}_{y},$ in $\bar{R}^{n}_{y}$ has coordinate values $(r_{y,x}a_{x})^{n}$ in $\bar{R}^{n}_{x}.$ This will be used in the following where some properties of geometry are discussed.

             \subsection{Line elements}
             Line elements are examples of local quantities that scale when referenced to different locations.  The usual four dimensional  expression, (sum over repeated indices) \begin{equation}\label{ds2y} ds^{2}(y)= g_{\mu,\nu}(y)dy^{\mu}dy^{\nu},\end{equation}is unchanged under the local availability of mathematics.  However, the metric tensor and  other factors are all elements of $\bar{R}_{y}.$

             The description of $ds^{2}(y),$ referenced to some location, $x$, is unchanged if scaling is absent. More specifically, the representation of $ds^{2}(y)$ at $x$, given by
             \begin{equation}\label{ds2yx}ds^{2}(y)_{x}=F_{x,y}g_{\mu,\nu}(y)dy^{\mu}dy^{\nu}= g_{\mu,\nu}(y)_{x}dy^{\mu}_{x}dy^{\nu}_{x}\end{equation} has the same value at $x$ as $ds^{2}(y)$ has at $y.$

             With scaling present, the representation of $ds^{2}(y)$  at $x$ becomes
             \begin{equation}\label{ds2Sc}ds^{2,Sc}(y)_{x}=e^{\Theta(y)_{x}-\Theta(x)}ds^{2}(y)_{x}.
             \end{equation} The scaling factor, Eq. \ref{rTxy}, and $F_{x,y},$ account for the change of reference point from $y$ to $x.$ If $x$ is in the region $L$, then use of Eq. \ref{Ty0} simplifies  Eq. \ref{ds2Sc} in that $\Theta(x)\simeq 0.$ This gives \begin{equation}\label{ds2Scy} ds^{2,Sc}(y)_{x}=e^{\Theta(y)_{x}}ds^{2}(y)_{x}.\end{equation}

             It is interesting to note that the dependence of the scaled line element on $y$ has the same form as  a conformal transformation \cite{conf} of the line element. The relation of the factor,  $exp(\Theta(y)_{x}),$ to conformal transformations and to conformal field theory \cite{Ginsparg} is not known at present.

             \subsection{Path Lengths}\label{PL}

              A good example of the effect of scaling on geometry is in the representation of path lengths. The length, $L(p)$, of a path $p$ extending from points $x$ to $y$ can be expressed as a line integral, $L(p)=\int_{p}|\vec{dp}|.$ If the path is parameterized by $s$ where $p(0)=x$ and $p(1)=y,$ then
              \begin{equation}\label{Lp}L(p)=\int_{0}^{1}|\nabla_{s}p|ds.\end{equation}Here $\nabla_{s}p$ is the gradient of $p$ at location $p(s).$ Also \begin{equation}\label{Nbp}|\nabla_{s}p|=(\nabla_{s} p\cdot\nabla_{s}p)^{1/2}.\end{equation}

              This representation of the path length holds in the usual setup with just one real number structure  for all points of $M$. However, Eq. \ref{Lp} is not valid under the assumption of local availability of mathematics.  The problem is that the integral represents the limit of a sum of real number integrand values in different real number structures, $\bar{R}_{p(s)}$. Addition is  defined within a real number structure.  It is not defined  for number values in different structures.

              This can be remedied by transfer of the integrands for different $s$  to a common reference point and then defining the integral as a limit of a sum of real number values in the real number structure at the reference point. If $x$ is chosen as the reference point, then parallel transfer of the integrands to the reference point before integration solves the problem.  The result is
              \begin{equation}\label{Lpx}L(p)_{x}=\int_{x,0}^{1}F_{x,p(s)}(|\nabla_{s}p|ds) =\int_{x,0}^{1}|\nabla_{s}p|_{x}ds_{x}.\end{equation} The subscripts, $x,$ indicate that the integral is evaluated at $x$ with integrand values all in $\bar{R}_{x}.$

              The fact that $L(p)_{x}$ is the same number value as $L(p)$ in Eq. \ref{Lp} shows that local availability of mathematics, without scaling, has no effect on path lengths. This examples joins others that show the same result:  local availability of mathematics without scaling does not affect theoretical description of physical and geometric properties.

              In the presence of scaling, the path length integral becomes\begin{equation}\label{LpScx} \begin{array}{l} L^{\Theta}(p)_{x}=\int_{x,0}^{1}e^{\Theta(p(s))_{x}-\Theta(x)} F_{x,p(s)}(|\nabla_{s}p|ds) \\\\\hspace{1cm}=\int_{x,0}^{1}e^{\Theta(p(s))_{x}-\Theta(x)} |\nabla_{s}p|_{x}ds_{x}.\end{array}\end{equation}

              The expression for the scaled path length can also be written as \begin{equation}\label{LScpx}
              L^{\Theta}(p)_{x}=e^{-\Theta(x)}\int_{x,0}^{1}e^{\Theta(p(s))_{x}}|\nabla_{s}p|_{x}ds_{x}.
              \end{equation} This shows clearly the difference between external scaling and internal scaling. The factor $\exp(\Theta(x))$ is an external scaling factor as it is outside of the mathematical operation, (integration over $p$ in $M$).  It corrects for  scaling at the reference point.   For a local quantity that does not involve integrals or derivatives over space or time, such as the line element, external scaling can always be removed by transferring the reference point to the location of the quantity.  This is shown by Eq. \ref{ds2Sc} if the reference point is changed to $y.$

              Internal scaling refers to the presence of scaling factors that are inside some mathematical operation and cannot be moved outside. The factor $\exp(\Theta(p(s))_{x})$ in Eq. \ref{LScpx} is an example of internal scaling as the factor depends on the integration variable and cannot be moved outside the integral. Also internal scaling factors cannot be removed by change of reference points.

              The location of $x$ at an endpoint of $p$ is arbitrary.  Any other point $z$ in $M$ can be chosen as a reference point. In this case Eq. \ref{LScpx} becomes \begin{equation}\label{LScpz}
              L^{\Theta}(p)_{z}=e^{-\Theta(z)}\int_{z,0}^{1}e^{\Theta(p(s))_{z}}|\nabla_{s}p|_{z}ds_{z}.\end{equation} The endpoints of $p$ are still at $x$ and $y$.

              The effect of scaling depends on the the location of the path $p$ in the universe.  If the path with its endpoints is contained in the local region, $L$ that can be populated by observers, then $\Theta(p(s))_{x}-\Theta(x)\simeq 0,$  and there is no scaling for the path.  Note that no scaling is a special case of scaling with the scaling factor equal to $1$ everywhere. If parts of the path are cosmological (the path is very long, or some or all of $p$ is outside $L$), then scaling would be present. The amount of scaling clearly depends  on the properties of $\Theta$.

              \subsection{Distances}\label{D}
              As might be expected, the presence of number scaling affects  distances between points.  With scaling present, the distance between two points, $x$ and $y$, is obtained by variation of   $L^{\Theta}(p)_{z}$, Eq. \ref{LScpz}, over the paths, $p$, and setting the result equal to $0.$  The resulting Euler Lagrange equation is \cite{BenNOVA}\begin{equation} \label{ELEU}\frac{\partial \Theta(p(s))}{\partial p_{\mu}}|\nabla_{s}p|-\frac{d}{ds}(\Theta(p(s))) \frac{\partial |\nabla_{s}p|}{\partial (\partial_{\mu,s}p)}=\frac{d}{ds}\frac{\partial |\nabla_{s}p|}{\partial (\partial_{\mu,s}p)}.\end{equation} Here $\partial_{\mu,s}p=dp_{\mu}/ds.$ The length of the path $p$ satisfying this equation is a minimum and is the distance between $x$ and $y.$

              Scaling introduces two $\Theta$ dependent terms into the equation. These are shown on the left hand side of Eq. \ref{ELEU}.  If $\Theta$ is constant, then both terms on the left hand side of the equation are zero and one obtains the usual equation for  distances as geodesics.

              \subsection{Time dependent $\Theta$}\label{TDT}

              So far,  descriptions of the effect of the field, $\Theta$ on geometry have been limited to general aspects, such as the effects on line elements and path lengths. No particular dependence of $\Theta$ on space or time has been assumed, other than the requirement that it be essentially constant for all points in the observer occupiable  region, $L$. Specific examples  are useful to further understand how the field affects physics and geometry.

               Time dependent scaling fields are interesting examples. Let $z=t,\textbf{z}$ be a space time location in $L$ of a potential observer. Events at all points, $x=s,\mathbf{x}$ in the past light cone of  $z$ are observable at $z.$ For any $\textbf{x}$, $x$ is in the past light cone of $z$ if $s\leq t-(|\mathbf{x}-\mathbf{z}|)/c.$ It is on the light cone of $z$ if \begin{equation}\label{stltcn}s=t-\frac{|\mathbf{x}-\mathbf{z}|}{c}.
              \end{equation}

              As a hypothetical example, assume that $\Theta(x)$ depends only on time and is independent of  space locations, $\Theta(t,\mathbf{x})=\Theta(t).$ Let the time $t$ be the present cosmological time, about $14\times 10^{9}$ years. Also use Eq. \ref{Ty0} to set $\Theta(t)=0.$ For events on the past light cone at distance $\mathbf{x}$ from $\mathbf{z}$, the reference space point in $L$, the scaling factor is
              \begin{equation}\label{esz}e^{\Theta(s)_{z}}=e^{\Theta(t-\frac{|\mathbf{z}-\mathbf{x}|}{c})_{z}}.
              \end{equation}

              Assume that $\Theta(s)_{z}$ decreases from $0$ at $s=t$ towards $-\infty$ as $s\rightarrow 0$, the time of the big bang. The scaling factor decreases from $1$ at time $t$ to $0$ at time $0$. It follows that all physical and mathematical quantities at time $s$, scaled to the present, approach $0$ as $s\rightarrow 0.$ As an example, distances between all space points approach $0.$ This is like the big bang in that all of space is crunched into a point.

              However it is also the case that energy densities of any cosmic dust, fluid, or matter,  also approach $0$ as $s$ approaches $0$. Not only this but the magnitudes of all physical quantities approach $0$ as $s\rightarrow 0.$  This is quite different from the accepted description of the big bang where quantities such as  energy densities approach $\infty.$

              If the time dependence of $\Theta$ is different in that $\Theta(s)\rightarrow \infty$ as $s\rightarrow 0,$ then  the magnitudes of all physical quantities close to $14\times 10^{9}$  lightyears distant become infinite.  This includes distances between points, energy densities, and other quantities.

              This is a good illustration of the fact that scaling affects mathematical quantities. Scaling factors are independent of which physical quantity, if any, is being considered and the magnitude of the quantity. Scaling factors depend only on the location, $x,$ of the quantity, e.g. energy density at $x$, and the location, $z$, of the reference point. Mathematical representations at point $x$ of physical quantities all change by the same scale factor when either $x$ or the reference point is changed. This corresponds to external scaling, described earlier in subsection \ref{PL}.

              \subsection{Black and white scaling holes}\label{BWSH}

              Other examples of the possible effects of the $\Theta$ field are illustrated by black and white scaling holes \cite{BenNOVA}.  Let $x_{0}$ be a point in three dimensional space. Let $\Theta$ be a time independent scaling field that is spherically symmetric about $x_{0}$. Then $\Theta(x)$ is the same for all points $x$ on a sphere of radius $r$ centered on $x_{0}.$

              Let $x$ be a reference point at an unscaled distance $r$ from $x_{0}$ and $y$ be a point on the radius vector between $x_{0}$ and $x.$ One is interested in determining the scaled distance from $x$ to $y$ as $y$ moves along the radius from $x$ to $x_{0}.$ For black scaling holes, $\Theta(y)\rightarrow\infty$ as $y\rightarrow x_{0}.$  For white scaling holes, $\Theta(y)\rightarrow -\infty$ as $y\rightarrow x_{0}.$

               A good description of examples is to give a specific dependence of $\Theta(y)$ as $y$ moves along the radius from $x$ to $x_{0}$ This can be done by letting $\Theta(y)$ depend on the unscaled distance, $s=|y-x_{0}|,$ between $y$ and $x_{0}.$ This is expressed here by an abuse of notation and writing $\Theta(y)=\Theta(s).$ Then the scaled distance from $x$ to $y$, referenced to $x$, is given by \begin{equation}\label{LsTh}L(y)^{\Theta}_{x}=L(s)^{\Theta}_{x}= \int_{x,0}^{r-s}e^{\Theta(r-u)-\Theta(r)}du.\end{equation} The integral is along the radius from $x$ to $y$ and $u$ is the  unscaled distance  between $x$ and any point $z$  on the radius between $x$ and $y.$

\begin{figure}[h!]\begin{center}\vspace{.1cm}\rotatebox{270}{\resizebox{150pt}{150pt} {\includegraphics[120pt,250pt][420pt,550pt]{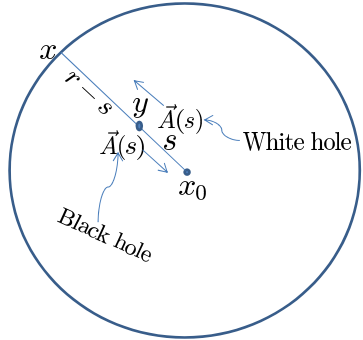}}}\end{center} \caption{Representations of the geometric parameters for black and white holes. The point $x$ on the circle whose points are all at an unscaled distance, $r,$ from the center at $x_{0},$ is a reference point. The point, $y,$ is at a distance, $s,$ from $x.$  The $\Theta$ gradient vectors, $\vec{A}(s),$ with arrows show the directions of the gradient of $\Theta$ for black and white holes}\label{SPIE32}\end{figure}

              Figure \ref{SPIE32} is a two dimensional slice through the sphere, including the center, that shows the relationship of the points. The circle is the locus of points equidistant from $x_{0}$ with the value of $\Theta$ the same for all points on the circle.

              As a specific example  let the $s$ dependence of $\Theta(s)$ be given by \begin{equation} \label{Ths}\Theta(s)=\frac{K}{s}.\end{equation}Using this in Eq. \ref{LsTh} gives \begin{equation}\label{LsuTh}L(y)^{\Theta}_{x}= \int_{x,0}^{r-s}\exp(\frac{K}{r-u} -\frac{K}{r})du.\end{equation} Replacement of $u$ by the dimensionless variable $z=u/r$ gives \begin{equation}\label{LszTh}L(y)^{\Theta}_{x}= r\int_{x,0}^{w}\exp(\frac{K}{r} (\frac{1}{1-z}-1))dz.\end{equation} The factor $r$ gives the  integral the dimension of length. $w=1-s/r$ is the fractional distance (distance divided by $r$) from $x$ to $y.$

              The effect of the $\Theta$ on the scaled length can be seen by choosing $x$ so that $r=1.$ For black holes set $K=r=1.$ The effect of scaling on the distance between $x$ and $y$ as a function of $w$ can be seen by plotting the integral in Eq. \ref{LszTh} as a function of $w$ as it ranges from $0$ to $1.$ The reference point remains at $x.$

              The results for a scaling black hole are shown in Figure \ref{SPIE33}.\footnote{The evaluation of the integrals and plots for this and the next figures are obtained online from www.rechneronline.de.} Both the scaled and unscaled distances from $x$ to $y$ are shown as a function of the fractional unscaled distance, $(r-s)/r=1-s$  from $x$ to $y$. The curves show these distances as $y$ moves from $x$ to $x_{0}.$\begin{figure}[h!]\begin{center}\vspace{2.5cm}\rotatebox{270}{\resizebox{150pt}{150pt} {\includegraphics[200pt,250pt][500pt,550pt]{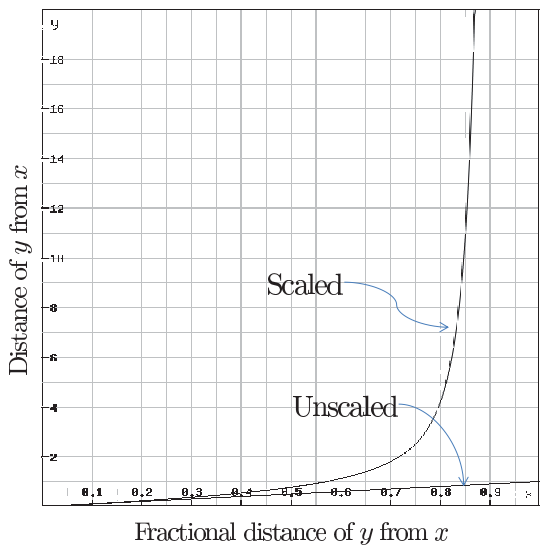}}}\end{center} \caption{Scaled and unscaled distances from a reference point, $x,$ to $y$ where $y$ is any point on the radius from $x$ to $x_{0}$. The distances are plotted as a function of the fractional unscaled distance, $1-s,$  of $y$ from $x.$ The vector $\vec{A}(y)=d\Theta(y)/dy$ is directed towards $x_{0}.$}\label{SPIE33}\end{figure}

              The curve for the scaled distance shows dramatically that the scaled distance from $x$ to $y$ increases rapidly to infinity as $y$ approaches $x_{0}.$ If $y$ is about $85\%$ of the distance for $x$ to $x_{0}$, then the scaled distance from $x$ to $y$ is about $10/.85=12$ times the unscaled distance from $x$ to $y.$

              The reason this is called a scaling black hole is that, if the path from $x$ to $x_{0}$ is parameterised as a function of $t$ as $\gamma(t)$  with $\gamma(0)=x$ and $\gamma(1)=x_{0}$, then the scaled distance as a function of $t$ is given by \begin{equation}\label{LvecTt} L(\gamma)^{\Theta}(t)_{x} =r\int_{0,x}^{t} e^{\Theta(\gamma(s))_{x}-\Theta(\gamma(0))}ds =r\int_{0,x}^{t} \exp(\frac{K}{r}(\frac{1}{1-s}-1))ds.\end{equation} If $t$ is the time and $\gamma(t)$ describes the motion of a  particle along the path, then the scaled speed of the particle towards $x_{0},$ referenced to $x$, is \begin{equation}\label{vSc} \frac{d(L(\gamma)^{\Theta}(t)_{x})}{dt}=\exp(\frac{K}{r}(\frac{1}{1-t}-1))_{x}.\end{equation} This increases exponentially to infinity as the particle approaches $x_{0}$ at $t=1.$

              The results for a scaling white hole are obtained by changing the sign of $K$ in Eq. \ref{LszTh}.
              In this case the scaled distance from $x$ to $y$ is given by \begin{equation}\label{LszThw}L(y)^{\Theta}_{x}= r\int_{x,0}^{w}\exp(\frac{-K}{r} (\frac{1}{1-z}-1))dz.\end{equation} Setting $K/r=1$ and $r=1$, as before gives \begin{equation} \label{LszThw1}L(y)^{\Theta}_{x}= \int_{x,0}^{w}\exp(-\frac{1}{1-z}+1))dz.\end{equation}

\begin{figure}[h]\begin{center}\vspace{2cm}\rotatebox{270}{\resizebox{150pt}{150pt} {\includegraphics[170pt,250pt][470pt,550pt]{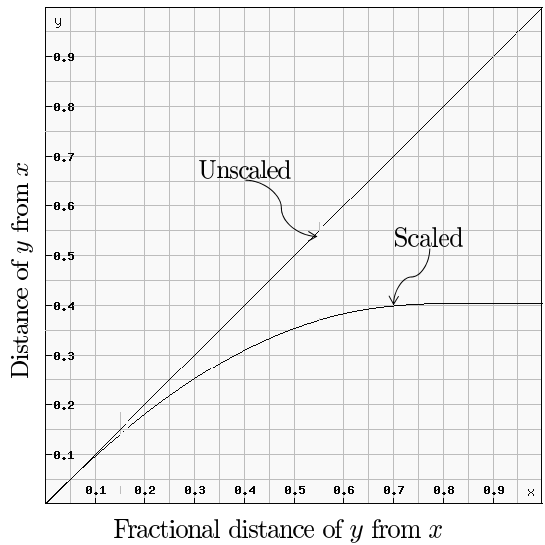}}}\end{center} \caption{Scaled and unscaled distances from a reference point, $x,$ to $y$ where $y$ is any point on the radius from $x$ to $x_{0}$. The distances are plotted as a function of the fractional unscaled distance, $1-s,$  of $y$ from $x.$ The vector $\vec{A}(y)=d\Theta(y)/dy$ is directed away from $x_{0}.$}\label{SPIE34}\end{figure}

               A plot of the scaled length from $x$ to $y$ as a function of the fractional distance of $y$ from $x$, as $y$ moves from $x$ to $x_{0},$ is given in Figure \ref{SPIE34}. The reference point is $x.$ In this case scaled distances are compressed relative to the unscaled distance.  The scaled distance from $x$ to $x_{0}$ is only $0.4$ of the unscaled distance.

              This is the opposite of the situation for positive $K$ where the scaled distance went to infinity as $y$ approached $x_{0}$. Here the scaled distance from $x$ to $x_{0}$ is a fraction of the unscaled distance. Using the same argument as was used for the black scaling hole, one sees that the speed of a particle at $\gamma(t),$ moving from $x$ towards $x_{0},$ approaches $0$ as $t\rightarrow 1.$

               These properties show why this case is called a white scaling hole.  The presence and outward direction of $\vec{A}(y)$ compress properties such as path lengths and particle velocities compared to their unscaled values. In this sense, $\vec{A}$ acts like an outward force  from $x_{0}$ on both geometric and physical properties of systems in that it pushes or compresses properties  outward from from  $x_{0}.$

               As might be expected, the effect of scaling on path lengths from $x$ out to more distant locations on the line from $x_{0}$ through $x$ is the opposite of the effect on locations between $x$ and $x_{0}.$ For black scaling holes, the $x$ referenced, scaled path lengths to points more distant from $x$ are compressed relative to those for unscaled path lengths. This occurs because the field $\Theta (y)$ increases as $y$ moves radially away from $x_{0}$ even beyond $x.$ The gradient field, $\vec{A}(y),$ is opposite to the direction of motion of $y.$

               For scaling white holes the $x$ referenced, scaled lengths to points $y$ out from $x$ are greater than the unscaled lengths. This occurs because the values of $\Theta(y)$ decrease as $y$ moves away from $x_{0}$ and the resultant gradient field $\vec{A}(y)$ is  directed outward from $x_{0}.$  This shows that for white scaling holes, moving outward, increases scaled path lengths relative to unscaled lengths. The same effect occurs moving inward for black scaling holes,  The reason is that the direction of $\vec{A}(y)$ is parallel to the direction of motion of $y$ in both cases.

               A similar situation holds for moving outward in scaling black holes and moving inward for scaling white holes. In this case the direction of motion of $y$ is antiparallel to the direction of $\vec{A}(y).$ More details on black and white scaling holes are given in \cite{BenNOVA}.

               \section{Discussion}

               Several points about the effects on physics and geometry of space and time dependent scaling of different number types  should be emphasized. Scaling relates  number structures of different types, and other mathematical systems based on numbers, at points $y,$ to a reference point $x.$ The scaling factor depends only on $x$ and $y.$ It is the same factor for all number values and for mathematical elements such as vectors in Hilbert spaces. So far, it has nothing to do with whether a number value is the value of a physical quantity or not. The scaling factor is the same for numerical values of all physical quantities.

               As a specific example, let $\lambda_{y}$ denote the wavelength of a photon at $y.$ This is a number value, in $\bar{R}_{y}$, for a length at $y.$ The scaled value of this length, as a number in $\bar{R}_{x},$ is $$\lambda^{r_{y,x}}_{x}=r_{y,x}F_{x,y}\lambda_{y}=r_{y,x}\lambda_{x}.$$ Here $\lambda_{x}$ is the same value in $\bar{R}_{x}$ as $\lambda_{y}$ is in $\bar{R}_{y}.$

               The same formula holds if $\lambda$ is replaced by $\gamma_{y}$, the frequency of the photon. In this case $\gamma^{r_{y,x}}_{x}=r_{y,x}\gamma_{x}.$  The scaling factor is the same as for the frequency. Similarly the speed of light, $c_{y},$ at $y$ corresponds to the scaled speed of light, $r_{y,x}c_{x},$ at $x.$  This shows that even fundamental physical constants are subject to the same scaling as are other physical quantities and pure numbers.

               These relations show an interesting property of scaling, that equations are invariant under transfer from $y$ to $x,$ even with scaling present. For example the relation between wavelength and frequency is given by $\lambda=c/\gamma.$ The local availability of mathematics requires that both sides of this equation be number values in the same real number structure.  One cannot have $\lambda$ a value in $\bar{R}_{y},$ and $c$ and $\gamma$ values in $\bar{R}_{x}$.

               The equation relating wavelength and frequency at $y$ is given by \begin{equation}\label{wfy}
                \lambda_{y}=\frac{c_{y}}{\gamma_{y}}\mbox{}_{y}.\end{equation} The subscripts $y$ indicate that the number values and division operation are in $\bar{R}_{y}.$ The scaled representation of this equation in $\bar{R}_{x}$ can be written in an intermediate form,\begin{equation}\label{wrfy}
                \lambda^{r_{y,x}}_{x}=(c^{r_{y,x}}_{x})\div^{r_{y,x}}_{x}(\gamma^{r_{y,x}}_{x}).\end{equation} This is an equation in the representation shown in Eq. \ref{Rxr}.  The three factors and the division operation all include scaling from $y$ to $x.$

                Use of the scaling of number values and operations shown in Eq. \ref{Rrx} give the scaled representation of Eq. \ref{wfy} on $\bar{R}_{x}$ as $$r_{y,x}\lambda_{x}=(r_{y,x}c_{x}) r_{y,x}\div_{x}(r_{y,x}\gamma_{x})$$ or\begin{equation}\label{lxcx}\lambda_{x} =c_{x}\div_{x}\gamma_{x}.\end{equation}  This is the same equation at $x$ as Eq. \ref{wfy} is at $y.$ Also the number values of the factors in this equation are the same in $\bar{R}_{x}$ as the number values of the factors in Eq.\ref{wfy} are in $\bar{R}_{y}.$

                This shows that equations are invariant under change of reference points, even with scaling present. This is good because it follows that theoretical predictions in physics, expressed as equations, are the same at all points, even with scaling present.

                The same argument holds for equations containing terms with space or time integrals or derivatives.  The equations are invariant under change of reference points, but the scaling factors inside  the integral or in the derivative remain.

                The invariance of equations under reference point change with number scaling is quite different from the change of scale and scale invariance as usually understood in physics \cite{Scale}. One difference is that scale factors in physics are not usually treated as being dependent on space and/or time locations.  Here the number scaling factors are dependent on space and time locations.

                Another difference is that scaling in physics is applied to the magnitudes of physical quantities only.  It is not applied to the operations such as multiplication or division that combine quantities. If $a$ is the magnitude of a physical quantity or constant and $\lambda$ a scale factor then $a^{2}$ scales as $(\lambda a)^{2}=\lambda^{2}a^{2}.$ Here if $a_{y}$ is the magnitude of a physical quantity at $y$ and $\lambda$ is the scaling factor from $x$ to $y,$ then the magnitude of the square, $a_{y}^{2}$  at  location $x$ is $(\lambda a_{x})\times/\lambda (\lambda a_{x})=\lambda a_{x}^{2}.$  Here $a_{x}$ is the same value at $x$ as $a_{y}$ is at $y.$

                The lambda factor in the denominator accounts for the scaling of the multiplication operation, under  transfer from $y$ to $x.$ This inclusion of scaling of the operations in number structures, and other mathematical systems based on numbers, is the source of the invariance of equations under change of reference points.

                This work, and earlier work, towards a theory of mathematics and physics together requires that one use a specific model of mathematical systems.  This was done here with the choice of representing mathematical systems of each type as structures \cite{Barwise,Keisler} that satisfy a set of axioms relevant to the system type being considered.   A possible connection to physics appears here in the assumed existence of a scalar boson field in gauge theories that represents the space time dependent  scaling of mathematical structure.

                It is clear that much more work needs to be done to connect these ideas more closely to physics.  At present it is not known if physics makes use of the scalar boson field for number scaling, or even if the local availability of mathematics, with or without scaling, has observable experimental consequences for physics.

                Even though there is much work to be done, it is hoped that this work is a real first step to developing a coherent theory of mathematics and physics together. The locality of mathematics, the freedom of choice of number systems and other  mathematical systems based on numbers, and the use of space time dependent number scaling have been seen to have many interesting consequences.

             \section*{Acknowledgement}
            This work was supported by the U.S. Department of Energy,
            Office of Nuclear Physics, under Contract No.
            DE-AC02-06CH11357.


\begin{thebibliography}{99}

               \bibitem{Wigner}
             Wigner, E.,  "The unreasonable effectiveness of mathematics in the natural sciences," Commum. Pure and Applied Math. {\bf 13}, 001 (1960), Reprinted in   "Symmetries and Reflections",
            Indiana Univ. Press, Bloomington IN, pp. 222-237.

             \bibitem{Omnes}
           Omnes, R.,  "Wigner's "Unreasonable Effectiveness of
           Mathematics", Revisited," Foundations of Physics,
           \textbf{41}, 1729-1739, (2011).

           \bibitem{Plotnitsky}
           Plotnisky, A., "On the reasonable and unreasonable effectiveness of mathematics in classical and quantum physics," Foundations of Physics, {bf 41}, 466-491, (2011).

           \bibitem{Hamming}
            Hamming, R. W., "The unreasonable effectiveness of mathemati8cs," Amer. Math Monthly, {\bf 87},  (1980).

               \bibitem{Tegmark}
            Tegmark, M., "The mathematical universe," Found. Phys., \textbf{38}, 101-150, (2008).


            \bibitem{BenCTPM1}
            Benioff, P., "Towards a Coherent Theory of Physics and Mathematics," Foundations of Physics,\textbf{32}, 989-1029, (2002);
            arXiv:quant-ph/0201093.

             \bibitem{BenCTPM2}
             Benioff, P., "Towards a Coherent Theory of Physics and Mathematics: The Theory-Experiment Connection."  Foundations of Physics,\textbf{35}, 1825-1856, (2005);  arXiv:quant-ph/0403209.

              \bibitem{Barwise}
            Barwise, J., "An Introduction to First Order Logic," in
            [Handbook of Mathematical Logic]                                                                    , J. Barwise, Ed.
            North-Holland Publishing Co. New York, 1977. pp 5-46.

            \bibitem{Keisler}
            Keisler, H. J.,  "Fundamentals of Model Theory," in
            [Handbook of Mathematical Logic], J. Barwise, Ed.
            North-Holland Publishing Co. New York, (1977). pp 47-104.

            \bibitem{rational}
             Weir, A. J., [Lebesgue Integration and Measure],
            Cambridge University Press, New York, NY, (1973), P. 12.


            \bibitem{complex}
             Shoenfield, J., [Mathematical Logic], Addison Weseley
            Publishing Co. Inc. Reading Ma, (1967), p. 86; Wikipedia:
            Complex Numbers.

            \bibitem{Kadison}
           Kadison, R. and  Ringrose, J.  [Fundamentals of
           the Theory of Operator Algebras: Elementary theory],
           Academic Press, New York, (1983), Chap 2.

           \bibitem{Lloyd}
            Lloyd, S., "The quantum geometric limit"," ArXiv:1206:6559.


            \bibitem{Mack}
             Mack, G., "Physical principles, geometrical aspects, and locality properties of gauge field theories," Fortshritte der Physik, \textbf{29}, 135
            (1981).

            \bibitem{BenNOVA}
               Benioff, P., "Gauge theory extension to include number scaling by boson field: Effects on some aspects of physics and geometry," To appear as chapter in [Recent Developments in Bosons Research], Nova publishing Co., (2013), arXiv:1211.3381

             \bibitem{BenSPIE2}
            Benioff, P., "Local availability of mathematics and number scaling: effects on  quantum physics," in Quantum Information and Computation X, Donkor, E.; Pirich, A.; Brandt, H., Eds.;
             Proceedings of SPIE, Vol. 8400; SPIE: Bellingham, WA, 2012, 84000T;  arXiv:1205.0200.


             \bibitem{BenINTECH}
            Benioff, P., "Effects on Quantum Physics of the Local Availability of Mathematics and Space Time Dependent Scaling Factors for Number Systems," [Advances in Quantum Theory], Ion I. Cotaescu (Ed.),  InTech, (2012), Available online at: http://www.intechopen.com/articles/show/title/effects-on
            -quantum-physics-of-the-local-availability-of
            -mathematics-and-space-time-dependent-scaling; arXiv:1110.1388.


             \bibitem{real}
             Randolph, J., [Basic Real and Abstract Analysis],
            Academic Press, Inc. New York, NY, (1968), P. 26.

            \bibitem{Yang}
             Yang, C. N. and  Mills, R. L., "Conservation of Isotopic Spin and Isotopic Gauge Invariance," Phys. Rev., \textbf{96}, 191-195,  (1954).



             \bibitem{Montvay}
             Montvay I. and  M\"{u}nster, G., [Quantum Fields on a
            Lattice], Cambridge Monographs on Mathematical Physics,
            Cambridge University Press, UK, (1994).

            \bibitem{Rothe}
             Rothe, H. J., [Lattics Gauge Theories, an Introduction], 3rd Edition,
            Lecture Notes in Physics, Vol. 74, World Scientific, Singapore, Chap 2. (2005).

            \bibitem{Cheng}
            Cheng, T. P. and  Li, L. F., [Gauge Theory of Elementary  Particle Physics], Oxford University Press, Oxford, UK, (1984), Chapter 8.

           \bibitem{conf}
             Tong, D., "Lectures on string theory",  arXiv:0908.0333, Chapter 4.

            \bibitem{AMSci}
           H. A. Smith, "Alone in the Universe", American Scientist, \textbf{99},  No. 4, p. 320, (2011).


           \bibitem{Ginsparg}
            Ginsparg, P., "Applied conformal field theory," arXiv:hep-th/9108028, (1988).

           \bibitem{Scale}
           Wikipedia: Scale Invariance.


            \end{thebibliography}
            \end{document}